\documentstyle[11pt,
epsfig]{article}
\input{psfig.sty}
\newcommand{\mut}{\tilde{\mu}}
\newcommand{\dr}{\Delta R}
\newcommand{\rv}{\vec{r}}

\newcommand{\pha}{\phi_1}
\newcommand{\phit}{\tilde{\phi}}

\newcommand{\phb}{\phi_2}

\newcommand{\lx}{\lambda}
\newcommand{\Lx}{\Lambda}
\newcommand{\ex}{\epsilon}

\newcommand{\sx}{\sigma}

\newcommand{\mt}{\tilde{m}^2}

\newcommand{\kx}{\kappa}

\newcommand{\be}{\begin{equation}}
\newcommand{\ee}{\end{equation}}
\newcommand{\een}{\end{subequations}}
\newcommand{\ben}{\begin{subequations}}
\newcommand{\beq}{\begin{eqnarray}}
\newcommand{\eeq}{\end{eqnarray}}

\def \lta {\mathrel{\vcenter
     {\hbox{$<$}\nointerlineskip\hbox{$\sim$}}}}
\def \gta {\mathrel{\vcenter
     {\hbox{$>$}\nointerlineskip\hbox{$\sim$}}}}

\begin{document}
%\twocolumn[
\centerline{\bf July 2001 \hfill hep-ph/0109080
} \vspace{3.5cm}

\centerline{\Large\bf Large Gauged Q Balls} \vspace{2.cm}

\centerline{ {\large \bf K.N. Anagnostopoulos$^{a}$}\footnote {e-mail
address : konstant@physics.uoc.gr}{\large ,} {\large \bf
M. Axenides$^{b}$}\footnote {e-mail address :
axenides@mail.demokritos.gr }{\large ,} {\large \bf E.G.
Floratos$^{b,c}$}\footnote {e-mail address :
manolis@mail.demokritos.gr }{\large ,} {\large and} {\large\bf N.
Tetradis$^{a,c}$}\footnote {e-mail address : tetradis@physics.uoc.gr}
} \vspace{0.3cm} \centerline{\em a) Department of Physics, University
of Crete, GR-71003 Heraklion, Greece}
\vspace{0.2cm} \centerline{\em b) Institure of Nuclear Physics,
N.C.R.P.S. Demokritos, GR-15310 Athens, Greece} \vspace{0.2cm}
\centerline{\em c) Nuclear and Particle Physics Sector, University of
Athens, GR-15771 Athens, Greece} \vspace{3.cm}

\centerline{\large\bf Abstract}
\begin{quote}\large\indent
We study Q-balls associated with local $U(1)$ symmetries.  Such
Q-balls are expected to become unstable for large values of their
charge because of the repulsion mediated by the gauge force.  We
consider the possibility that the repulsion is eliminated through the
presence in the interior of the Q-ball of fermions with charge
opposite to that of the scalar condensate. Another possibility is that
two scalar condensates of opposite charge form in the interior.  We
demonstrate that both these scenaria can lead to the existence of
classically stable, large, gauged Q-balls.  We present numerical
solutions, as well as an analytical treatment of the ``thin-wall''
limit.

\end{quote}
\vspace{1cm}
%]

\newpage

\section{Introduction}
Non-topological solitons named Q-balls can appear in scalar field theories
with $U(1)$ symmetries \cite{coleman}.
(For a review of the early literature see ref. \cite{leepang}.)
These objects can be viewed as coherent states of the scalar field
with fixed total $U(1)$ charge. The case of global $U(1)$
symmetries has attracted much attention. The reason is the presence
of such symmetries in the Standard Model, related to baryonic or
leptonic charge. In supersymmetric extensions of the Standard Model,
the scalar superpartners of baryons or leptons can form coherent
states with fixed baryon or lepton number, making the existence
of Q-balls possible. Their properties \cite{existence,iiro},
cosmological origin
\cite{formation} and experimental implications \cite{experiment}
have been the subject of several recent studies.
Non-abelian global symmetries can also lead to the existence of
Q-balls \cite{nonabelian}.

We are interested in the less popular case of
Q-balls resulting from local $U(1)$ symmetries \cite{axen,local}.
Such Q-balls become unstable for large values of their charge
because of the repulsion mediated by the gauge force.
However, small Q-balls can still exist.
A possibility that has not been considered before is that
the repulsion is eliminated through the presence in the interior
of the Q-ball of fermions with charge opposite to that of
the scalar condensate.
The fermions must carry
an additional conserved quantum number that prevents their
annihilation against the condensate. This scenario can
lead to the existence of large Q-balls. The fermion gas may also be replaced
by another scalar condensate, of opposite charge to the first, such that
the interior of the Q-ball remains neutral.

In the following we discuss in detail the above scenaria in the context
of a toy model. We show that arbitrarily large Q-balls can exist
and examine the constraints imposed on the parameters by the requirement
of classical stability.
%We also discuss charge evaporation from the surface and
%demonstrate that it induces only small modifications to the basic picture.

\section{Small Gauged Q-Balls}
For completeness we summarize briefly the basic properties of gauged Q-balls
in a toy model (see ref. \cite{local} for the details).
We consider a complex scalar field
$\phi(\rv,t)=f(\rv,t)\,\exp(-i\,\theta(\rv,t))/
\sqrt{2}$, coupled to an Abelian gauge field
$A^\mu$. The Lagrangian density is
\be
{\cal L} = \frac{1}{2} \partial_\mu f\partial^\mu f+ \frac{1}{2} f^2
\left( \partial_\mu\theta-e A_\mu \right)^2 - U(f)
-\frac{1}{4} F_{\mu\nu} F^{\mu\nu}.
\label{toylag} \ee
The total $U(1)$ charge of a particular field configuration is
\be
Q_\phi=\int d^3\rv f^2 \left( \dot\theta -e A_0 \right).
\label{charge0} \ee
Without loss of generality we assume $e,Q_\phi \geq 0$ in the following.
The value $e=0$ leads
to decoupling of the scalar from the gauge field.

We consider a spherically symmetric ansatz
that neglects the spatial components of the
gauge field $A_i=0,~i=1,2,3$
and assumes $\theta = \omega t$ \cite{coleman}. The component
$A_0$ of the gauge field corresponds to the
electrostatic potential that is responsible for the repulsive force
destabilizing the Q-ball.
The equations of motion for the fields are
\beq
f''+\frac{2}{r}f'+fg^2-\frac{dU(f)}{df}&=&0
\label{eom1} \\
g''+\frac{2}{r}g'-e^2f^2g&=&0,
\label{eom2} \eeq
with $r=|\rv|$, $g(r)=\omega-eA_0(r)$, and primes denoting derivatives
with respect to $r$.
The total charge and energy are
\beq
Q_\phi &=& \int dV \, \rho_\phi=4 \pi \int  r^2\, dr \, f^2 g
\label{charge} \\
E&=& \int dV \, \ex=4 \pi \int r^2\, dr \left[
\frac{1}{2} f'^2 + \frac{1}{2e^2}g'^2 +\frac{1}{2} f^2 g^2
+U(f)\right].
%= \frac{1}{2} \omega Q_\phi + \frac{4\pi}{3} R^3 U(F).
\label{energy}
\eeq

The Q-ball solution of the equations of motion
involves
an almost constant non-zero scalar field $f(r)=F$
in the interior of the Q-ball,
which moves quickly to zero (the vacuum value) at the surface.
We are interested in the limit in which
the radius $R$ of the Q-ball is much larger than the thickness of its
surface and the total charge can be big.
In this limit and for
$eFR \ll 1$,
the energy can be expressed as \cite{local}
\be
E= Q_\phi \left[ \frac{2\,U(F)}{F^2} \right]^{1/2}
\left[ 1+ \frac{C^{2/3}}{5} \right]=
 Q_\phi \left[ \frac{2\,U(F)}{F^2} \right]^{1/2}
+\frac{3e^2Q_\phi^2}{20\pi R},
\label{approx2} \ee
with
\beq
R&=& \left[
\frac{3Q_\phi}{4\pi F\sqrt{2\, U(F)}} \right]^{1/3}
\left[ 1+ \frac{C^{2/3}}{45} \right]
\label{approx1a} \\
C&=&\frac{3e^3Q_\phi}{4\pi}
 \sqrt{\frac{F^4}{2U(F)}}.
\label{approx1b} \eeq

The ratio $E/Q_\phi$ increases with $Q_\phi$ because of the
presence of the electrostatic term
in eq. (\ref{approx2}).
This means that large Q-balls are unstable and
tend to evaporate
scalar particles from their surface
in order to increase their binding energy.
For small $Q_\phi$ the above expressions are not applicable. Numerical
solutions indicate that
the ratio $E/Q_\phi$ becomes large again because of the contribution
from the field derivative terms that we neglected.
Therefore, there is a value
$\left( Q_\phi \right)_{min}$ for which $E/Q_\phi$ is minimized.
Classical stability requires that
$\left( E/Q_\phi \right)_{min} < d^2U(0)/df^2$ (assuming that the
absolute minimum of the potential is at $f=0$), so that
the Q-ball does not disintegrate into scalar particles of unit charge.

\section{Large Gauged Q-Balls with fermions}
It is clear from the above discussion that gauged
Q-balls with very large $Q_\phi$ become unstable because of electrostatic
repulsion. One possibility that could remedy this problem is that
fermions with charge
opposite to that of the scalar background neutralize the
electrostatic field and eliminate the repulsion.
%These fermions carry
%an additional conserved quantum number that forbids their annihilation
%against the background.
A model that realizes this scenario has a Lagrangian density
\be
{\cal L} = \frac{1}{2} \partial_\mu f\partial^\mu f+ \frac{1}{2}
f^2 \left( \partial_\mu\theta-e A_\mu \right)^2 - U(f) +i
\bar{\psi}_\alpha \gamma^\mu \left( \partial_\mu +ie' A_\mu
\right) \psi^\alpha -\frac{1}{4} F_{\mu\nu} F^{\mu\nu}
\label{toyflag} \ee
In the absence of Yukawa couplings, the scalar
and fermionic fields carry independent conserved $U(1)$ charges. A
linear combination of these charges is gauged while the orthogonal
one remains global. We assume that there are $N$ fermionic degrees
of freedom, labelled by $\alpha=1...N$, with gauge coupling $e'$
and negligible mass. Realistic scenaria could involve condensates
of electrically charged mesonic fields, with characteristic scales
for their potentials ${\cal O}$(100 MeV -- 1 GeV), and Higgs or
squark fields, with characteristic scales ${\cal O}$(100 GeV -- 1
TeV). In all these cases, neutralizing fermions, such as
electrons, can be considered effectively massless.

The equation of motion (\ref{eom1}) is not altered by the presence
of fermions, but eq. (\ref{eom2}) becomes
\be
g''+\frac{2}{r}g'-e^2f^2g- e e' \psi^\dagger_\alpha \psi^\alpha=0.
\label{eom3} \ee
Instead of solving the
Dirac equation,  we approximate the fermions as
a non-interacting Fermi gas with position dependent density. This is
the Thomas-Fermi approximation \cite{schiff}.
The fermionic $U(1)$ charge and energy density are
\be
\left\langle \psi^\dagger_\alpha  \psi^\alpha \right\rangle
= N \rho_\psi = N \frac{k_F^3}{3 \pi^2},
~~~~~~~~~~~~~~
\left\langle \psi^\dagger_\alpha
\left( -i \vec{\alpha}\cdot \vec{\nabla} \right) \psi^\alpha \right\rangle
=N \ex_\psi = N \frac{k_F^4}{4 \pi^2},
\label{erhopsi} \ee
in terms of the Fermi momentum $k_F$.
The Dirac equation for a fermion near the Fermi surface results in the
expression
\be
\mu_\psi  = k_F(r) + e' A_0(r)
= k_F(r) + \frac{e'}{e} \left (\omega -g(r) \right).
\label{mupsi} \ee
We see that $\mu_\psi$ can be interpreted as the chemical potential,
i.e. the energy cost in
order to add an extra fermion on the top of the Fermi sea. The
fermions rearrange themselves so that $\mu_\psi$ is position
independent. It is convenient to define the gauge-invariant
chemical potential
\be
{\tilde{\mu}} = \mu - \frac{e'}{e} \omega
= k_F(r) - \frac{e'}{e} g(r).
\label{mutpsi} \ee

The total
energy is now given by
\be
E= 4 \pi \int r^2\, dr \left[
\frac{1}{2} f'^2 + \frac{1}{2e^2}g'^2 +\frac{1}{2} f^2 g^2
+U(f) + N \ex_\psi +
\left(
\vec{E} \cdot \vec{\nabla} A_0 + N e' \rho_\psi A_0 + e \rho_\phi A_0
\right)
\right],
\label{energyf} \ee
where $\rho_\phi$ is given by eq. (\ref{charge}), $\rho_\psi$ by the first of
eqs. (\ref{erhopsi})  and $\vec{E}$
is the electric field.
The equations of motion can be obtained by minimizing the energy under
constant scalar and fermionic charge. This can be achieved through the
use of Lagrange multipliers $\omega$ and $\mu_\psi$. Minimization of
$E - \omega \int \rho_\phi
 \,dV - \mu_\psi N \int \rho_\psi\, dV$ with respect to
$A_0$, $f$ and $k_F$ results in eqs. (\ref{eom3}), (\ref{eom1}) and
(\ref{mupsi}), respectively.
Finally, the quantity in parentheses in the rhs of eq. (\ref{energy})
vanishes through the application of Gauss' law, eq. (\ref{eom3}).

\begin{figure}[t]
%\vspace{1.cm}
\centerline{\psfig{figure=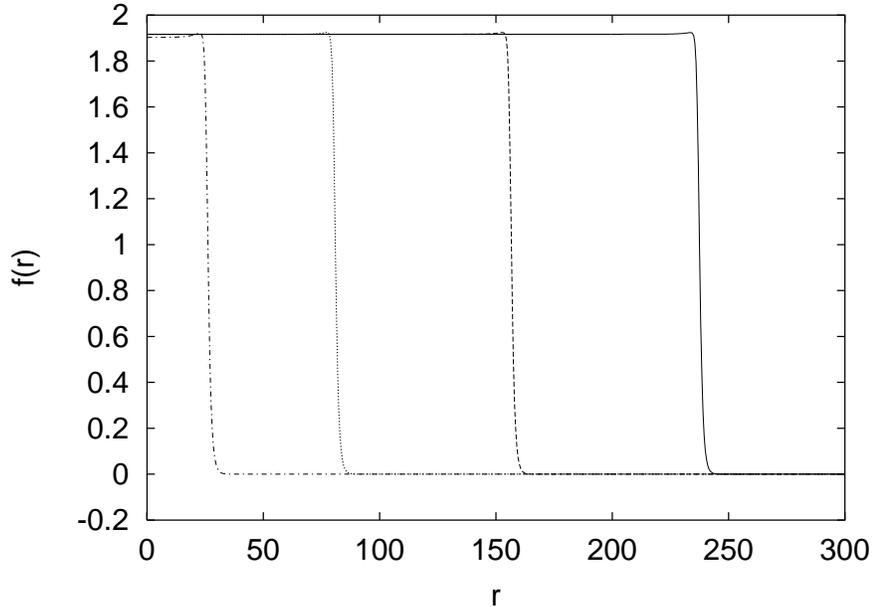,width=12cm}}
\caption{The magnitude of the
scalar field $f$ as a function of the radial distance $r$ for
Q-balls of increasing size.
}
\label{fig1}
\end{figure}

The above considerations provide a simple method for the
determination of
the properties of large Q-balls. We are interested in the ``thin-wall''
limit, in which the effects of the surface of the Q-ball may be neglected.
(A more careful discussion of the validity of this approximation is given
in the next section.) In this limit, the total energy is given by
\be
E = \left( \frac{1}{2}f^2 g^2 + U(f) + N \frac{k_F^4}{4 \pi^2} \right) V,
\label{enthin} \ee
with $V$ the volume of the Q-ball.
The charges of the scalar condensate and the fermions are
\be
Q_\phi = f^2 g V, ~~~~~~~~~~~~~~~~~~~ Q_\psi = N \frac{k_F^3}{3
\pi^2} V. \label{charges} \ee In terms of the constant scalar and
fermion charges $Q_\phi$ and $Q_\psi$ the total energy to be
minimized is given by:
\be
E= \frac{1}{2} \frac{Q_\phi^2}{f^2 V} + U(f) V +
\frac{(3 \pi^2)^{4/3}}{4 \pi^2 N^{1/3}} \,\,\frac{Q_\psi^{4/3}}{V^{1/3}}.
\label{enthin2} \ee

Minimization with respect to $f$ and use of the first of eqs. (\ref{charges})
gives
\be
fg^2 = \frac{dU(f)}{df}\equiv U^{\prime}. \label{con2} \ee This
relation could have been obtained by requiring that eq.
(\ref{eom1}) be satisfied for constant fields. The existence of
such a solution for eq. (\ref{eom3}) leads to
\be
e f^2 g = - e' N \frac{k_F^3}{3 \pi^2}. \label{con3} \ee This
implies
 \be
e'Q_\psi+eQ_\phi=0 \label{neutrality} \ee and guarantees the
electric neutrality of the interior of the Q-ball.

We can also obtain the equilibrium volume of our large fermion
Q-ball. It is given by \be V = \frac{Q_\phi}{ \sqrt{f^3
U^{\prime}}}\label{vol}\ee Minimization of eq. (\ref{enthin}) with
respect to $V$ results in the relation
\be
U(f)=\frac{1}{2}f^2 g^2 + \frac{1}{3} N \frac{k_F^4}{4 \pi^2}=
\frac{1}{2}f^2 g^2 + \frac{1}{3} \epsilon_\psi. \label{con1} \ee
It can be put in a more convenient equivalent form, for which the
scaling between $Q_\phi$ and $V$ is explicit. Expressed solely in
terms of $f$ it takes the form
\be
2U = f U^{\prime} + \frac{\left(3\pi^2 \right)^{4/3}}{6
\pi^2}\left| \frac{e}{e'}\right|^{4/3} \frac{1}{N^{1/3}}f^2 \left[
U^{\prime} \right]^{2/3}. \label{final}
 \ee

 Equations (\ref{con2}),(\ref{con3}) and (\ref{final})  uniquely determine the values
of $f$, $g$, $k_F$ in the interior of a large Q-ball. As a
consequence, $\mut$ can be also specified  through eq.
(\ref{mutpsi}). It is now obvious that the total energy scales
linearly with $Q_\phi$ for a given set of values for $f$ and $g$
\be
\frac{E}{Q_\phi} = \left[  \frac{1}{2}\sqrt{\frac{U^\prime}{f}} +
\frac{U}{\sqrt{f^3 U^\prime}} + \frac{\left(3\pi^2
\right)^{4/3}}{4\pi^2}\left| \frac{e}{e'}\right|^{4/3}
\frac{1}{N^{1/3}} \left(f^3 U^\prime\right)^{1/6}\right].
\label{Energy}\ee

For massless fermions, the stability condition $\min (E/Q_\phi) <
\sqrt{U''(0)}$ guarantees that a large gauged Q-ball cannot
disintegrate into a collection of free particles. One could also
consider the possibility that a $\phi$ particle is surrounded by a
``cloud'' of fermions (or the other way around), so that the
resulting ``atom'' is approximately neutral. A collection of such
states would probably be energetically favourable to a collection
of free particles, due to the electrostatic attraction. However,
for couplings $e,|e'| \lta 1$, the electrostatic binding energy is
expected to be much smaller than the mass of the free scalars,
similarly to the situation in normal atoms. For this reason, the
above relation gives a sufficiently accurate criterion for the
classical stability of Q-balls.

 In the limit $e \to 0$, eqs. (\ref{con2})--(\ref{con1}) give
$k_F=0$ and the well-known conditions for the existence of global
Q-balls are reproduced : $\omega^2= E^2/Q^2= 2U(f)/f^2=
U^{\prime}/f$ \cite{coleman}.

\begin{figure}[t]

%\vspace{1.cm}
\centerline{\psfig{figure=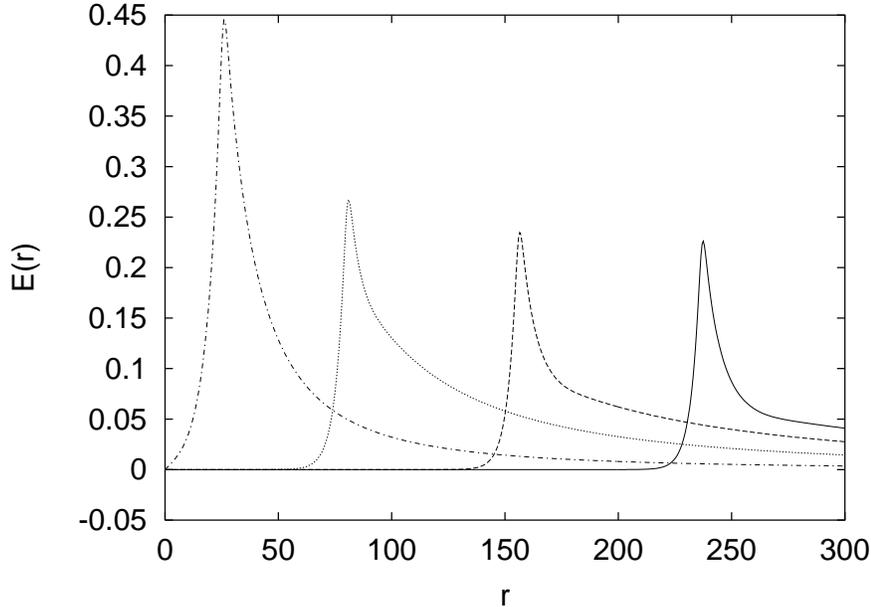,width=12cm}}
\caption{The electric
field $E$ as a function of the radial distance $r$ for
Q-balls of increasing size.
}
\label{fig2}
\end{figure}

\section{Numerical solutions}

In this section we present numerical solutions of the
equations of motion (\ref{eom3}), (\ref{eom1}) and
(\ref{mupsi}). The two differential equations require
four boundary conditions. We impose
$f'(r=0)=g'(r=0)=0$, so that there are no singularities at
the center of the Q-balls. We also impose
$f(r=\infty)=0$ and $g'(r=\infty)=0$, so that the solutions
outside the Q-balls
correspond to the normal vacuum.
We use a potential of the form
$U(f)=f^2/2-f^4/4+\lx^2 f^6/6$, in order to make comparisons with
the results of ref. \cite{local}. For the same reason
we choose $\lx^2=0.2$ and $e=0.1$. We assume that there are $N=10$
fermionic species of unit charge $e'=-0.2$. A large number of
species results in a small fermionic kinetic
energy that helps to keep the Q-balls classically stable.
Moreover, values $N={\cal O}(10)$ are typical of realistic
theories, such as the MSSM.
All dimensionful quantities are
considered to be renormalized with respect to the mass term
in the potential (set equal to 1).

Q-ball solutions of various sizes are obtained by
fixing the value of $\mu$ and varying $\omega$.
The chemical potential $\mu$ is assumed to be negative. The reason
is apparent through eq. (\ref{mupsi}). If we would like to interpret
$A_0$ as the electrostatic potential, we must choose a gauge
such that $A_0(r) \propto r^{-1}$ for large $r$.
By taking $\mu$ negative, we expect that $k_F$ will become 0 at a finite
radial distance. We assume that there are no fermions at larger distances, so
that eq. (\ref{mupsi}) is inapplicable.
Instead we impose $k_F=0$ in eq. (\ref{eom3}), which results in the expected
behaviour for $A_0(r)$. Positive values of $\mu$ would result in a non-zero
fermionic density at arbitrary distances from the center of the Q-ball.

In figs. 1--3 we present a series of Q-ball solutions of increasing
size. In fig. 1 we plot the scalar field $f$ as a function of the
radial distance from the center of the Q-ball. We observe that $f(r)$
behaves as a step function to a very good approximation,
even for fairly small Q-balls. In fig. 2 we depict the magnitude of
the electric field for the same solutions. The smallest Q-ball has
the strongest electric field. The field vanishes at the center for
symmetry reasons, but quickly grows with $r$. For large enough $r$ it
falls $\propto r^{-2}$. For larger Q-balls the electric field is zero in
the interior, because of the cancellation of the charge of the
scalar field by that of the fermionic gas. The electric field is non-zero
near the surface, while it falls again $\propto r^{-2}$ for large
$r$.

\begin{figure}[t]
%\vspace{1.cm}
\centerline{\psfig{figure=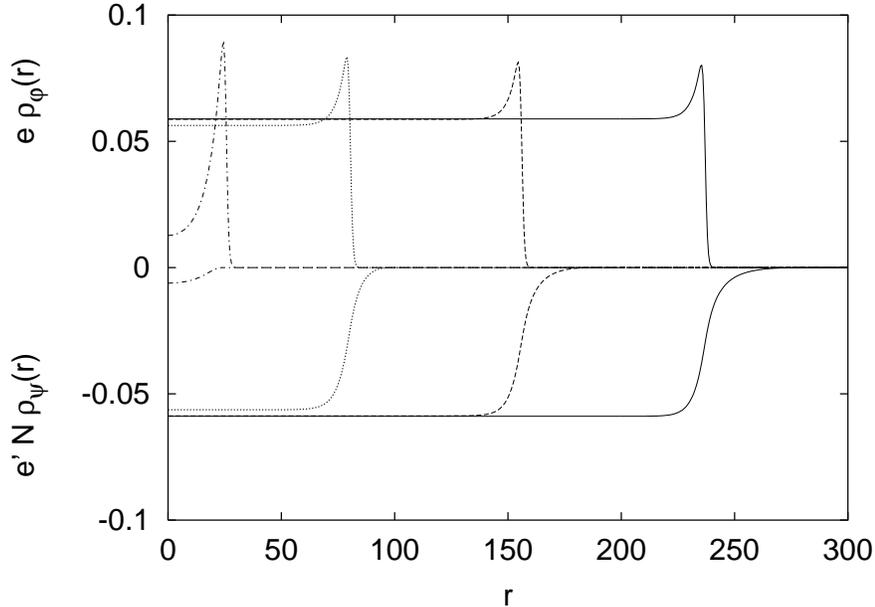,width=12cm}}
\caption{The charge densities of
the scalar condensate and the fermionic gas
as a function of the radial distance $r$ for
Q-balls of increasing size.
}
\label{fig3}
\end{figure}

In fig. 3 we plot the scalar and fermionic charge densities as a function
of the radial distance. For the smallest Q-ball the fermions
are not capable to neutralize the interior. There is a mismatch between the
scalar and fermionic densities. Moreover, there is a large concentration
of scalar charge near the surface. This is a result of the electrostatic
repulsion that forces the positive unit charges to maximize the distance
between themselves. As the size of the Q-ball increases, the charge densities
in the interior become opposite to each other. For large Q-balls
their magnitude is independent of the radius.
This is the ``thin-wall'' limit we discussed in the previous section.
The values of $f$, $g$ and $k_F$ in the interior (and, therefore,
the charge densities) should be uniquely determined
by eqs. (\ref{con2})--(\ref{con1}).
We have checked that $f(r=0)$, $g(r=0)$ and $k_F(r=0)$ for the large
Q-ball solutions
depicted in figs. 1--3 satisfy eqs. (\ref{con2})--(\ref{con1}) with an
accuracy better than 1\%.

A particular question merits some discussion at this point. From figs. 1--3
one could infer naively
that the profile of the surface is constant for large Q-balls
and merely displaced at different radii $R$. This would mean that the electric
field is the same near the surface for all large Q-balls. Such a field can
only be produced if the net surface charge density is constant and the
net surface charge scales $Q_s \propto R^2$. One implication
would be that the electrostatic contribution to the total energy of the
system
$\sim Q_s^2/R \propto R^3$ would scale proportionally to the
volume. As a result, our assumption
that the surface effects are negligible in the ``thin-wall'' limit would
be invalid.
However, the numerical solutions do not confirm this picture.
The shape of the numerical solution varies slightly at the surface
even for large Q-balls. The electric field at the surface
becomes smaller for increasing radius, while the fermionic density is modified
appropriately. Numerically
we have not identified any residual surface effect. Moreover,
we expect that a more rigorous treatment of the fermionic cloud that surrounds
the Q-ball would support this conclusion. Our simple approximation of the
fermions as a non-interacting gas is adequate for the interior but very crude
near the surface. We expect
that, in a more careful treatment of the surface, a surrounding
fermionic cloud will neutralize completely the Q-ball
(similarly to the neutralization of atoms).
In this picture, the surface effects would be even less important in the
``thin-wall'' limit.

\begin{figure}[t]
%\vspace{1.cm}
\centerline{\psfig{figure=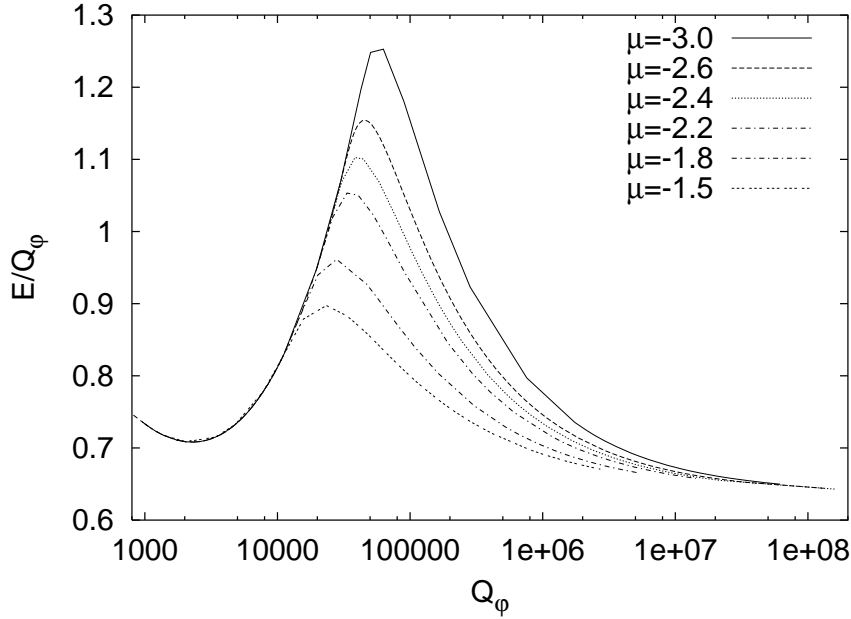,width=12cm}}
\caption{The energy to charge ratio $E/Q_\phi$ as a function of the
scalar charge $Q_\phi$ of the Q-ball.
}
\label{fig4}
\end{figure}

In fig. 4 we plot the energy to charge ratio as a function of
the charge $Q_\phi$ of the scalar condensate.
The fermionic charge $Q_\psi$ may differ substantially from $Q_\phi$ for
small Q-balls. As we have assumed that the fermions are massless and
normalized everything with respect to the scalar mass term,
the classical stability requirement is
$E/Q_\phi < 1$. In fig. 4 we observe a series of curves that correspond to
different (negative) values of $\mu$. The fermionic content of a
small Q-ball is controlled through $\mu$ and, for
the same $Q_\phi$, the various curves have
different ratios $Q_\psi/Q_\phi$.
In fig. 5 we plot $Q_\psi/Q_\phi$ as a function of $Q_\phi$
for the same range values of $\mu$ as in fig. 4.
We observe that $Q_\psi/Q_\phi$
tends to increase with decreasing $|\mu|$.
As the ratio $E/Q_\phi$ decreases for decreasing $|\mu|$,
we conclude that, for fixed $Q_\phi$,
the Q-balls become more stable by absorbing fermions
and increasing their fermionic content.
A limit to the process of fermion accretion
is set by the requirement of a positive chemical
potential, so that the fermions are bound to the Q-ball.

\begin{figure}[t]
%\vspace{1.cm}
\centerline{\psfig{figure=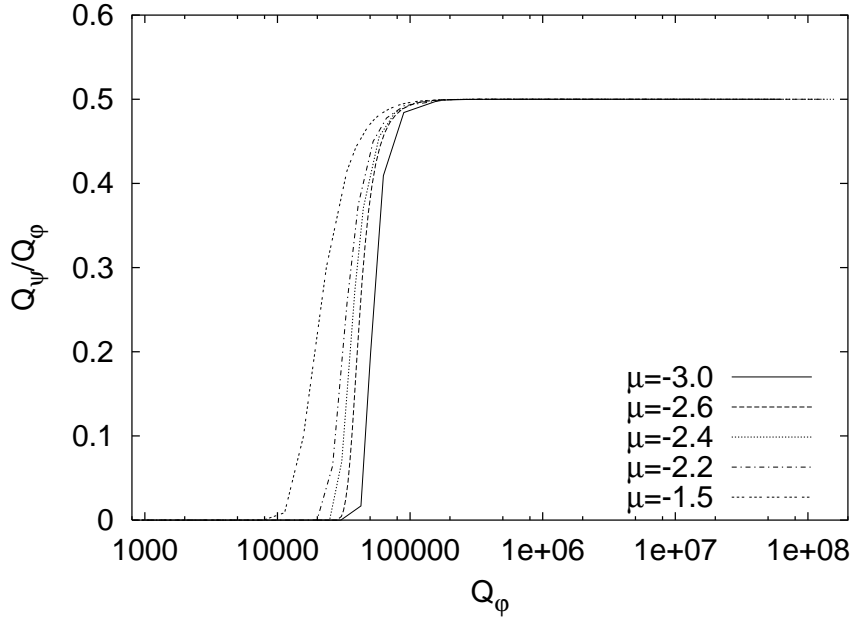,width=12cm}}
\caption{The ratio of scalar to fermionic charge
$Q_\psi/Q_\phi$ as a function of the
scalar charge $Q_\phi$ of the Q-ball.
}
\label{fig5}
\end{figure}

For $Q_\phi\lta 10^4$, the fermionic content becomes negligible
and we obtain the gauged Q-balls of ref. \cite{local}. For
$Q_\phi\lta 10^3$ the ratio $E/Q_\phi$ increases because of the
contribution of the derivative terms to the energy \cite{local}.
For $Q_\phi\gta 10^7$, $Q_\psi/Q_\phi=1$ and the energy to charge
ratio has a very weak dependence on $\mu$ and $Q_\phi$. In this
region the "thin-wall" approximation is valid. The gauge invariant
quantity $\mut$ of eq. (\ref{mutpsi}) is almost constant.
Asymptotically for $Q_\phi \to \infty$, the properties of the
Q-balls are completely determined by the values of $f$, $g$ and
$k_F$ in the interior as given by the solution of eqs.
(\ref{con2})--(\ref{con1}).

\begin{table}[b]
\begin{center}
\begin{tabular}{|l|l|l|l|l|l|l|l|}
\hline
& $N=1$ & $N=2$  & $N=4$  & $N=6$   & $N=10$ & $N=15$ & $N=20$ \\ \hline
$e'=-0.1$ & 1.954 & 1.654 & 1.403 & 1.277 & 1.135 & 1.036 & 0.972 \\ \hline
$e'=-0.2$ & 1.021 & 0.878 & 0.759 & 0.700 & 0.635 & 0.589 & 0.560 \\ \hline
$e'=-0.3$ & 0.724 & 0.633 & 0.559 & 0.522 & 0.481 & 0.453 & 0.435 \\ \hline
\end{tabular}
\caption{
The energy to charge ratio $E/Q_\phi$ in the ``thin-wall'' limit, for
various values of the parameters $N$ and $e'$.
}
\label{param}
\end{center}
\end{table}

For our choice of parameters, the biggest Q-balls are the most
stable. Moreover, as we discussed above, the stability is enhanced
by the absorption of fermions. These results suggest an efficient
accretion mechanism for large Q-balls with important astrophysical
implications \cite{starwreck}. For different parameters it is
possible that small Q-balls with $Q_\phi \sim 10^3$ and without
fermions become the most stable states. However, it is apparent
from fig. 4 that a barrier would still separate the large from the
small Q-balls. The decay of large Q-balls into smaller fragments
and free fermions would involve tunnelling and probably would
proceed at a very slow rate.

The dependence of the ratio $E/Q_\phi$ on $N$ and $e'$ in the ``thin-wall''
limit is given in table 1. The various values have been obtained
through the numerical solution of the algebraic system of equations
(\ref{con2})--(\ref{con1}) for $e=0.1$.
We observe that a small number $N$ of fermionic species with $|e'|=e$
results in a high energy to charge ratio and, therefore, unstable
Q-balls. This behaviour is caused by the big contribution from
the fermionic kinetic energy. Large values of $N$ permit the
distibution of the compensating charge among various
species, thus reducing the fermionic energy $\propto N^{-1/3}$.

\section{Large Gauged Q-Balls with two Scalar Condensates}

Another possibility is that two scalar condensates with opposite charges
form in the interior of a gauged Q-ball.
An appropriate Lagrangian density is
\be
{\cal L} = \frac{1}{2} \partial_\mu f\partial^\mu f + \frac{1}{2}
f^2 \left( \partial_\mu\theta_1-e A_\mu \right)^2 +\frac{1}{2}
\partial_\mu \chi \partial^\mu \chi + \frac{1}{2} \chi^2 \left(
\partial_\mu\theta_2-e' A_\mu \right)^2 - U(f,\chi) -\frac{1}{4}
F_{\mu\nu} F^{\mu\nu} \label{toyfxlag} \ee
The two scalar fields
carry independent conserved $U(1)$ charges, a linear combination
of which is gauged. We assume a time dependence for the two
condensates of the form: $\theta_1=\omega_1 t$, $\theta_2=\omega_2
t$. The resulting equations of motion are \beq
f''+\frac{2}{r}f'+f(\omega_1-e A_0)^2-\frac{\partial
U(f,\chi)}{\partial f}&=&0 \label{eomn1} \\
\chi''+\frac{2}{r}\chi'+\chi (\omega_2-e' A_0)^2 -\frac{\partial
U(f,\chi)}{\partial \chi}&=&0 \label{eomn2} \\
A_0''+\frac{2}{r}A_0'+ef^2 (\omega_1-e A_0)+e'\chi^2 (\omega_2-e'
A_0)&=&0. \label{eomn3} \eeq The energy of the system is given by
the expression \beq E= 4 \pi \int r^2\, dr &\Biggl\{& \frac{1}{2}
A_0'^2 +\frac{1}{2} f'^2 +\frac{1}{2} f^2 (\omega_1-e A_0)^2
+\frac{1}{2} \chi'^2 +\frac{1}{2} \chi^2 (\omega_2-e' A_0)^2
+U(f,\chi) \Biggr. \nonumber \\ &+& \Biggl. \left( \vec{E} \cdot
\vec{\nabla} A_0 + e f^2 (\omega_1-e A_0) A_0 +e' \chi^2
(\omega_2-e' A_0) A_0 \right) \Biggr\}. \label{energyfx} \eeq
Similarly to the discussion in section 3, the equations of motion
(\ref{eomn1})--(\ref{eomn3}) can be obtained by minimizing the
total energy under constant total charges of the two scalar
condensates. The expression in the second line of eq.
(\ref{energyfx}) vanishes through application of Gauss' law, eq.
(\ref{eomn3}).

The numerical solution of the three second-order differential
equations (\ref{eomn1})--(\ref{eomn3})
is more difficult than in the case of a scalar condensate with
compensating fermions. In that case we had to integrate two
second-order differential equations and an algebraic one.
Moreover, we expect a qualitative behaviour very similar to
the one studied in sections 3 and 4. For a large Q-ball to remain
classically stable, the net charge in its interior must be zero.
A possible mis-match at the surface could result in non-zero
electrostatic energy. However, in the ``thin-wall'' limit
this contribution is expected to become negligible.
For this reason, we limit our discussion to the analytical treatment
of the ``thin-wall'' limit, which is more useful for practical applications.

In this limit, the total energy is given by
\be
E = \left( \frac{1}{2}f^2 g_1^2 +\frac{1}{2}\chi^2 g_2^2 + U(f,\chi)
\right) V,
\label{enthinx} \ee
with $V$ the volume of the Q-ball and $g_1=\omega_1-e A_0$,
$g_2=\omega_2-e' A_0$.
The charges of the two scalar condensates are
\be
Q_\phi = f^2 g_1 V, ~~~~~~~~~~~~~~~~~~~ Q_\chi = \chi^2 g_2 V
\label{chargesx} \ee

They are taken to satisfy an electric charge neutrality condition
\be  e Q_{\phi} + e' Q_{\chi}=0 \label{neutral}\ee Keeping each of
them fixed means that we must minimize the quantity
\be
E= \frac{1}{2} \frac{Q_\phi^2}{f^2 V} +\frac{1}{2}
\frac{Q_\chi^2}{\chi^2 V} + U(f,\chi) V
 \label{enthin2x} \ee

Minimization with respect to $V$ results in the relation
\be
U(f,\chi)=\frac{1}{2}f^2 g_1^2+ \frac{1}{2}\chi^2 g_2^2.
\label{con1x} \ee Three more constraints can be obtained by
requiring that eqs. (\ref{eomn1})--(\ref{eomn3}) be satisfied for
constant fields. They are
 \beq fg_1^2 &=& \frac{\partial
U(f,\chi)}{\partial f}, \label{con2x} \\ \chi g_2^2 &=&
\frac{\partial U(f,\chi)}{\partial \chi}, \label{con3x} \\ e f^2
g_1 &+& e'\chi^2 g_2=0 \label{con4x} \eeq The first two could have
been obtained through the minimization of the total energy of eq.
(\ref{enthin2x}) with respect to $f$ and $\chi$. The last equation
guarantees the electric neutrality of the interior of the Q-ball.
The four equations (\ref{con1x})--(\ref{con4x}) uniquely determine
the gauge-invariant quantities $f$, $\chi$, $g_1$, $g_2$ in the
interior of a large Q-ball. As a consequence, $E$, $Q_\phi$,
$Q_\chi$ are also specified through eqs. (\ref{enthinx}),
(\ref{chargesx}). Similarly to the fermionic case, the fixed
charges $ Q_\phi, Q_\chi $ scale linearly with the volume for
fixed values of their interior field variables
$f,\chi,g_1,g_2$.One reason for this is the charge neutrality
condition in eq.(\ref{neutral}) which they satisfy. Hence the
total energy of the double condensate configuration scales
linearly with respect to the scalar charge $Q_{\phi} = |e'/e|
Q_{\chi}$. The classical stability condition becomes $\min(E) <
m_\phi Q_\phi + m_\chi Q_\chi$, with $m_\phi^2=\partial^2
U(0,0)/\partial \phi^2$, $m_\chi^2=\partial^2 U(0,0)/\partial
\chi^2$ the masses of the two scalars at the vacuum at
$\phi=\chi=0$.

\section{Conclusions}

The main emphasis in the studies of Q-balls has been on
theories with global $U(1)$ symmetries. Theories with local $U(1)$
symmetries can support Q-balls as well. However, in the absence of
a neutralizing mechanism, the electrostatic repulsion destabilizes the
Q-balls with significant charge.
In the main part of this paper
we pointed out that gauged Q-balls can be stabilized through the
neutralization of the scalar condensate by fermions of
opposite charge. The total energy is increased because of the kinetic energy
of the fermions. However, the resulting configuration can be
stable even for arbitrarily large charge of the scalar condensate.

>From a cosmological perspective, the neutralization of gauged
Q-balls is expected.
For example, one could envision the existence of electric Q-balls, that could
be produced during phase
transitions \cite{formation} when the Universe passes through
an electric-charge breaking vacuum \cite{zarikas}.
It seems likely that several fermionic species could be trapped within
the Q-ball during its formation. The ones with charge of similar sign to
the scalar condensate will be expelled, so that the resulting object
will remain approximately neutral.

The existence of large electric fields can lead to spontaneous
pair creation. The presence of a
strong electrostatic field at the surface of the Q-ball can
separate a virtual fermion-antifermion pair
and bring the particles on mass shell \cite{morris}.
The fermion will be attracted towards
the surface, while the antifermion will be expelled.
In the vacuum, the critical field strength is $E_{crit}=m_\psi^2/|e'|$.
Our assumption that the fermion mass is much smaller than the typical
scale of the potential of the scalar field implies that this mechanism
is very efficient. In the interior of
a Q-ball, the pair creation stops only when the fermionic energy levels
are populated
up to a Fermi momentum comparable to the scale of the scalar field potential.
It seems, therefore, likely that
large gauged Q-balls can be neutralized through this mechanism, instead of
disintegrating.

We mention that evaporation from the surface is possible if the
scalar field has decay channels into light species \cite{evaporate}.
In this case, simultaneous evaporation of the decay products and fermions
maintains the approximate neutrality of the Q-ball.

The tendency of gauged Q-balls to trap fermions in their interior could have
interesting experimental consequences.
Even though we concentrated on masslees fermions, heavy exotic species may
have found their way to the interior of gauged Q-balls. Thus the
discovery of a Q-ball may lead to the additional discovery
of the exotic species trapped in its interior. The fact that
the energy per charge of a Q-ball is reduced when its fermionic
content is increased
(up to neutralization) indicates an efficient accretion mechanism with
important astrophysical implications \cite{starwreck}.

%{\bf Discuss}

We also discussed the possibility of neutralization of a gauged
Q-ball through the presence of two scalar condensates of opposite
charge in its interior. In this case the formation of Q-balls seems less
likely than in the case of one scalar condensate with compensating
fermions.
For neutral Q-balls to be produced, two
condensates with the appropriate properties (values of $f$, $\chi$,
$\omega_1$, $\omega_2$) must be assumed to be generated dynamically after a
phase transition \cite{formation}. This should be more difficult than
the trapping of fermions from the thermal bath in the region with a
non-zero charged condensate.

Finally, we point out that the neutralization mechanism is expected to work
for general potentials of the scalar field. In particular, we expect it
to be applicable to the case of potentials with flat directions, such
as the ones appearing in supersymmetric extensions of the Standard Model.
In this case
the global Q-balls do not approach the ``thin-wall'' limit, even though
they can become very big, with energy that scales $E \propto Q^{3/4}$
\cite{dvali}. The gauged Q-balls with similar potentials cannot reach
large sizes, unless a neutralization mechanism (through trapping of
fermions for example) eliminates the electrostatic repulsion.

\paragraph{Acknowledgements:} We  would like to thank
L. Perivolaropoulos for helpful discussions.  The work of K.N.A. and
N.T. was supported by the European Commission under the RTN programs
HPRN--CT--2000--00122, HPRN--CT--2000--00131 and
HPRN--CT--2000--00148. The work of K.N.A. was also supported by the RTN
program HPRN--CT--1999-00161, a National Fellowship Foundation of
Greece (IKY) and the INTAS contract N 99 0590.

%\newpage

%\small

\end{document}